\documentclass[aps,prd,twocolumn,showpacs,superscriptaddress,nofootinbib]{revtex4-1}
\usepackage{graphicx}  
\usepackage{dcolumn}   
\usepackage{bm}        
\usepackage{amssymb}   
\usepackage{color}
\usepackage{aas_macros}

\newcommand{\msun}{\ensuremath{\mathrm{M}_\odot}}
\newcommand{\sv}[1]{\textcolor{black}{#1}}

\newcommand{\Event}{GW150914}
\newcommand{\Xmas}{GW151226}

\newcommand{\si}{\ensuremath{\sim}}
\newcommand{\mc}{\ensuremath{\mathcal{M}}}

\newcommand{\massrangeinj}{\ensuremath{[12,200]~\msun}}
\newcommand{\qmininj}{1/3}
\newcommand{\beq}{\begin{equation}}
\newcommand{\eeq}{\end{equation}}
\newcommand{\degg}{deg\ensuremath{^2}}

\begin{document}

\title{Parameter estimation for binary black holes with networks of third generation gravitational-wave detectors}
\author{Salvatore Vitale}
\author{Matthew Evans}
\affiliation{LIGO, Massachusetts Institute of Technology, Cambridge, Massachusetts 02139, USA}

\date{\today}
\begin{abstract}
The two binary black-hole (BBH) coalescences detected by LIGO,  \Event{} and \Xmas{}, were relatively nearby sources, with a redshift of \si{}0.1. As the sensitivity of Advanced LIGO and Virgo increases in the next few years, they will eventually detect \sv{stellar-mass} BBHs up to redshifts of \si1.
However, these are still relatively small distances compared with the size of the Universe,
 or with those encountered in most areas of astrophysics.
In order to study BBH during the epoch of reionization,
 or black holes born from population III stars,
 more sensitive instruments are needed.
Third-generation gravitational-wave detectors, such as the Einstein Telescope or the Cosmic Explorer are already in an advanced R\&D stage. These detectors will be roughly a factor of 10 more sensitive \sv{in strain} than the current generation,
 and be able to detect BBH mergers beyond a redshift of 20.
 In this paper we quantify the precision with which these new facilities will be
  able to estimate the parameters of stellar-mass, heavy, and intermediate-mass BBH
  as a function of their redshifts and the number of detectors.
We show that having only two detectors would result in relatively poor estimates
 of black hole intrinsic masses; a situation improved with three or four instruments.
Larger improvements are visible for the sky localization,
 although it is not yet clear whether BBHs are luminous in the electromagnetic or neutrino band.
The measurement of the spin parameters, on the other hand, do not improve significantly as more detectors are added to the network since redshift information are not required to measure spin.
\end{abstract}

\maketitle

\section{Introduction}

With the detection of the binary black holes (BBH) \Event{}~\cite{GW150914-DETECTION} and \Xmas{}~\cite{GW151226-DETECTION,O1BBH} the era of gravitational wave (GW) astrophysics has started.
The first two systems detected by the LIGO and Virgo collaborations had very different masses. \Event{} was made of two black holes of roughly $30$\msun\ each~\cite{GW150914-PARAMESTIM,O1BBH}, i.e. much more massive than known stellar-mass black holes~\cite{2010ApJ...725.1918O}. These large masses have been used to set constraints on the metallically of the progenitor stars~\cite{GW150914-ASTRO,2010ApJ...715L.138B}. At $14$~\msun{} and $7$~\msun{}, the masses of \Xmas{} were instead in the middle of the range of masses for known black holes (BHs)~\cite{O1BBH}. Very little could be said about the spins of either source~\cite{O1BBH}, mostly due to the lack of visible precession~\cite{2016arXiv160601210T,O1BBH}.

Although very different in their physical parameters, the two events had something in common: their luminosity distance, which was slightly more than $400$Mpc. Using the cosmology measured by the latest Planck results~\cite{2015arXiv150201589P}, this corresponds to a redshift of $\sim 0.09$~\cite{O1BBH}.

Over the next few years, existing ground-based GW detectors such as LIGO~\cite{TheLIGOScientific:2014jea,Harry:2010zz} and Virgo~\cite{AVirgo} will steadily increase their sensitivities~\cite{2016LRR....19....1A}. Once at design sensitivity, toward the end of this decade, they will be a factor of 10 more sensitive in strain than 1st generation GW detectors (initial LIGO and Virgo). Other detectors will join the network: KAGRA~\cite{2012CQGra..29l4007S} is being built in Japan, while the construction of LIGO India~\cite{Indigo} has been recently approved.
This network of second generation (advanced) detectors will be able to probe a significant volume, and detected heavy BBH up to redshifts of unity (with heavier and optimally oriented systems detectable up to $z\sim2$~\cite{VitaleDiff}).
A combination of better coatings, quantum squeezing and heavier test masses can add another factor of \si{}2 in (luminosity distance) range~\cite{2015PhRvD..91f2005M}, after which current facilities will saturate their potential.

New facilities (henceforth third-generation, or 3G, detectors) will be required to substantially increase the sensitivity beyond the advanced detectors.  These new detectors will allow us to explore the most remote corners of Universe, detect rare events, and explore phenomena which radiate GWs more weakly than
compact binary systems (e.g., core-collapse supernovae~\cite{2016PhRvD..93d2002G} and isolated neutron stars~\cite{Bejger2015}).

The Einstein Telescope~\cite{2010CQGra..27s4002P,2011CQGra..28i4013H} (ET) is a European proposal for an underground 3G detector.  Although its design is not yet precisely finalized, it should consist of 3 Michelson interferometers~\footnote{In this paper we will refer to the whole ET apparatus, made of three interferometers, as one detector.} with 10 Km long arms, and inter-arm angles of 60$^\circ$, arranged to form a triangle. The fact that three interferometers are used gives the ET more power in discriminating GW polarizations than an equivalent L-shaped detector~\cite{2009CQGra..26h5012F}. On the other hand, the fact that they are co-located strongly reduces the capabilities to localize GW sources on the sky. Finally, if built underground, it would have good sensitivity down to a few Hz, \sv{due to lower Newtonian noise~\cite{2015PhRvD..92b2001H}}, as opposed to the $\sim$10~Hz realistically achievable with above-ground detectors.  

Another possible way forward to 3G detectors is to keep orthogonal arms, but significantly increase their length. Cosmic Explorer (CE)~\cite{2016arXiv160708697A,2014arXiv1410.0612D} is a proposed ground-based 40 Km L-shaped detector. Intense R\&D will be necessary, and is already ongoing, to ensure that all the major known sources of noise can be dealt with. These include quantum noise, Newtonian noise and coating thermal noise~\cite{2016arXiv160708697A}.

3G detectors will have three main, related, advantages over existing instruments.
First, they will allow for extremely frequent detections of common systems (such as BBH) and will dramatically increase the probability of detecting rarer or weaker events, such as core-collapse supernovae (SNe).
Second, they will make a much larger fraction of the Universe accessible to GW observation.
As we will see below, 3G detectors will be sensitive to BBH up to redshifts of more than 20, well within the epoch of reionization.
Detection of extremely high redshift BBH might shed light on population III stars and on primordial black holes.
Finally, events at small redshifts (below unity) would be detectable with extremely large signal-to-noise ratios (SNR). For example, a CE class facility would detect systems similar to \Event{} with SNR of the order of a thousand. 

Several authors have analyzed many of the scientific goals that would be achievable with 3G detectors. However these works mostly covered the ET~\cite{2012CQGra..29l4013S} and focussed on binary neutron stars (BNS),
 since those were thought to be the most common sources of GWs in the Universe.
 Examples include tests of general relativity~\cite{2010ApJ...725.1918O,Mishra:2010tp},
  measurement of cosmological parameters~\cite{2015arXiv150606590D,2011PhRvD..83b3005Z,2010CQGra..27u5006S,2012PhRvD..86b3502T}, and measurement of the equation of state of neutron stars~\cite{2015arXiv150606590D,2012PhRvL.108i1101M}.

In this paper we consider the capabilities of 3G networks to characterize BBHs. We show that having a network of 3G detectors will be fundamental to extracting key parameters of the sources, such as their masses.
Since the most likely future detector network is currently unknown,
 we consider several hypothetical networks of 3G observatories, made of two, three or four sites.
Then, in order to determine the precision with which BBH
 parameters can be estimated by each hypothetical network,
 we simulate astrophysical populations of BBH.

One immediate result is that in order to accurately measure the masses of BHs in binaries  we must have a network with 3 or more detectors. 
This happens because what is measured with GW are the redshifted masses,
 from which the intrinsic masses are derived.
The is done by obtaining the redshift from the luminosity distance measurement
 and assuming a model cosmology.
Since the distance information is encoded in both polarizations of the GW signal, and strongly correlated with the inclination angle, more than two detectors are required. 
In what follows we show how the estimation of the (intrinsic)
 mass parameters improves by a factor of 2 (for nearby events) to a factor of several several
 if four 3G detectors are used instead of two.
We also show that, unsurprisingly,
 the same effect does not apply to the dimensionless spins, since they do not get redshifted.

This paper is structured as follows. In Sec.~\ref{Sec.Networks} we describe the networks configurations we considered. In Sec.~\ref{Sec.Injections} we describe the simulated BBH events and make some considerations on the role of multiple detectors. In Sec.~\ref{Sec.IntrMass} we expand on some details about the mass measurement. The main results are reported in Sec.~\ref{Sec.Results}, while some caveats are listed in Sec.~\ref{Sec.Caveats}. Conclusions and future work are summarized in Sec.~\ref{Sec.Conclusions}.

\section{Networks}\label{Sec.Networks}

In this study, we considered five different 3G network configurations, listed here in increasing number of instruments:
\newline
\vspace{-10pt}

\begin{itemize}
\item \textbf{LC} Two CE detectors, one in the location of LIGO Livingston and one in China (see table~\ref{Tab.IFOs} for details).
\item \textbf{LE} One CE in Livingston and one ET in Europe. 
\item \textbf{LAE} Two CE instruments, one in Livingston and the other in Australia, plus one ET in Europe.
\item \textbf{LCAE} Three CE instruments, one in Livingston one in Australia and one in China, plus one ET in Europe.
\item \textbf{LCAI} Four CE instruments, one in Livingston, one in Australia, one in China and one in India.
\end{itemize}

We do not consider hybrid networks with 2G and 3G instruments, since the much larger sensitivity of those latter would make 2G instruments superfluous.

We stress that the coordinates and orientations we used, Tab.~\ref{Tab.IFOs} are not meant to represent actual locations. In particular we did not check for geographical constraints or seismic noise levels.
These kind of detailed studies will of course have to be performed before a site is selected. However, for the purposes of this study the exact locations do not matter, and approximate positions (up to a few thousands kilometers) are enough. 

\begin{figure}[htb]
\includegraphics[width=0.9\columnwidth]{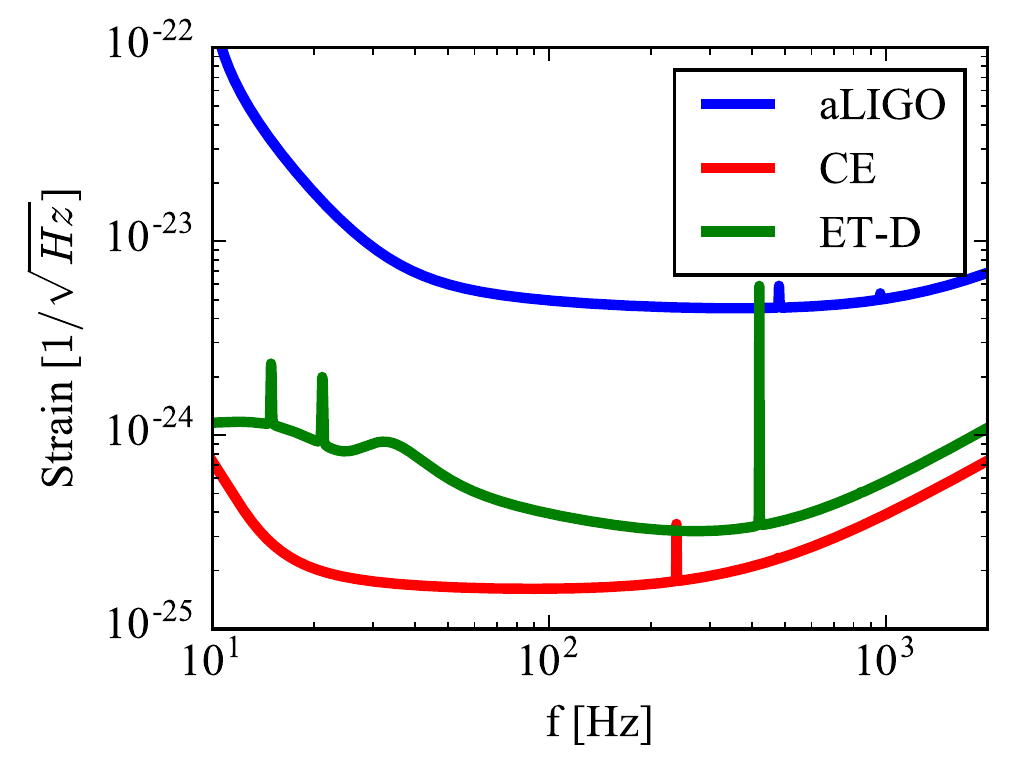}
\caption{The amplitude spectral density of the ET-D and CE, compared with the Advanced LIGO design. The curves can be downloaded from Ref.~\cite{2016arXiv160708697A}}
\label{Fig.Noises}
\end{figure}

For each detector we generated simulated gaussian noise using the power spectral densities for the ET-D and CE configurations given in refs.~\cite{2016arXiv160708697A,2011CQGra..28i4013H}, and shown in Fig.~\ref{Fig.Noises}.
While the Einstein Telescope could have good sensitivity down to a few Hertz, we started the analysis for all interferometers at the same frequency of 10Hz. However this will not change our conclusions, since those mostly depend on the geometry of the networks and not on the details of each instrument.

\begin{table}[htb]
\centering
\begin{tabular}{|c|c|c|c|c|} \hline
~~~~~  & ~Longitude~ & ~Latitude~  & ~Orientation~ & ~Type~ \\
  \hline
L &    -1.58       &      0.533       &  2.83    &     CE  \\  \hline
C &   1.82        &  0.67        &     1.57        &  CE       \\  \hline
I &     1.34      &   0.34       &      0.57      &     CE     \\  \hline
E &      0.182     &    0.76      &     0.34        &     ET      \\  \hline
A &      2.02     &    -0.55      &       0      &       CE \\\hline
\end{tabular}
\caption{The coordinates of the interferometers used in this study. Orientation is the smallest angle made by any of the arms and the local north direction. All angles are in radians. The last column reports the type on instrument; Cosmic Explorer (CE)~\cite{2016arXiv160708697A} or Einstein Telescope (ET)~\cite{2010CQGra..27s4002P}.}
\label{Tab.IFOs}
\end{table}

\section{Simulated BBH sources}\label{Sec.Injections}

In this section we describe the generation of the simulated BBH sources for each network.

We assumed that the intrinsic, or source-frame, total masses are uniform in the range \massrangeinj{} with a minimum mass ratio of \qmininj{}~\footnote{We define the mass ratio in the range $[0,1]$.}, to be consistent with the range of validity of the waveform family we used (see below). 
We notice that recent work suggests that in both globular cluster and galactic field evolutions the mass ratios of BBH will typically be in the range we are considering~\cite{2016Natur.534..512B,2016PhRvD..93h4029R}.
The lower limit of the component mass range is due to the evidence that stellar mass black holes have masses above $\sim6$\msun. The upper limit is somewhat arbitrary, since no observational evidence exists of intermediate-mass black holes.

Spins were uniform in magnitude in the range $[0,0.98]$ and randomly oriented on the unit sphere.

\begin{figure}[htb]
\includegraphics[width=0.9\columnwidth]{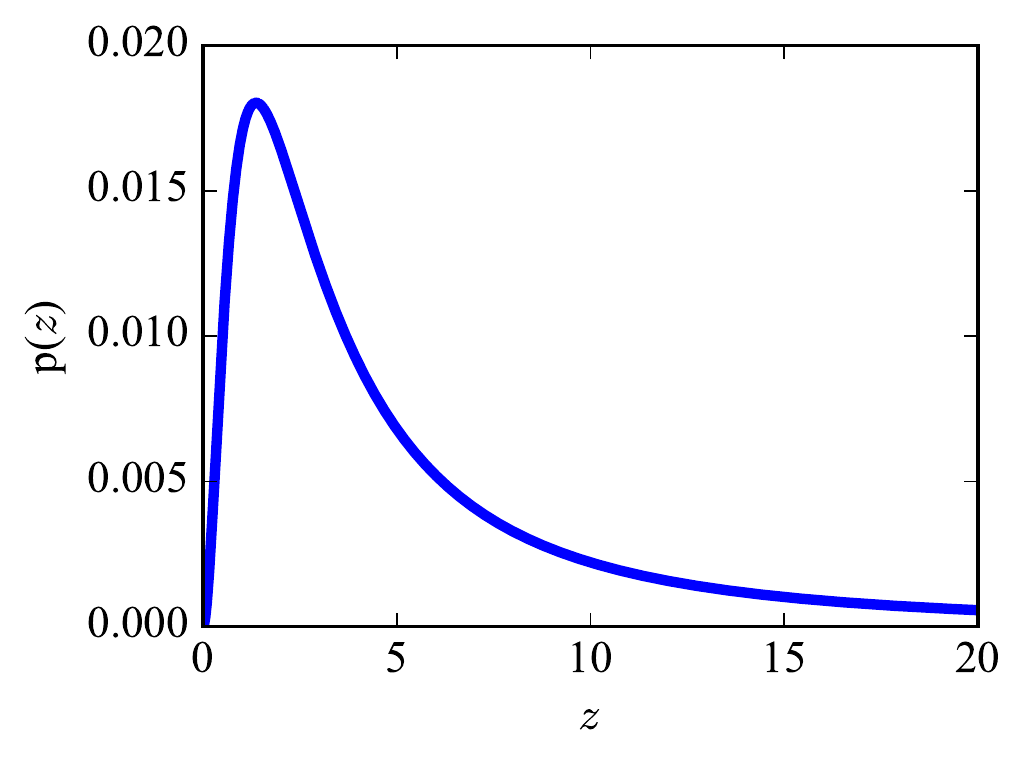}
\caption{The redshift distribution used to generate the simulated events. This curve is also used as prior for the parameter estimation algorithm.}
\label{Fig.RedshiftPrior}
\end{figure}

The redshifts were uniform in comoving volume, assuming a standard $\Lambda$CDM cosmology~\cite{2015arXiv150201589P}~\footnote{We used $\Omega_M = 0.3065$, $\Omega_\Lambda = 0.6935$ and  $H_0 = 67.90$ km s$^{-1}$ Mpc$^{-1}$}, in the range $z \in [0,20]$. We thus assumed the merger rate is not a strong function of the redshift, which of course is only a rough approximation.
However the main goal of this paper is not as much to report astrophysical uncertainties, as to show how those uncertainties depend on the GW network used. We thus assumed this to be a sufficient working hypothesis. If our readers have a particular merger rate in mind, they will be able to use our figures in the range of redshift where they expect most sources.
The redshift distribution we used is shown in Fig.~\ref{Fig.RedshiftPrior}.

For each set of proposed parameters randomly generated from the distributions described above, we calculate the SNR it would produce in the network under consideration, and only keep the source if the SNR is in the range $[10,600]$. However we notice that this requirement was seldom used,  i.e., for all networks most of the proposed sources had a SNR inside this range.

In this paper we do not deal with confusion noise and detectability of sources.
Work exists in the context of ET for binary neutron stars~\cite{PhysRevD.86.122001,2009PhRvD..79f2002R,2016PhRvD..93b4018M} where it has been shown that even overlapping events can be detected very efficiently (since the overlap in time needs not to correspond to an overlap in frequency). We will assume that the same is true for BBH and only use the SNR as a probe for detectability. A full mock data challenge will be put in place for a network of 3G detectors to fully support this assumption, which is beyond the scope of this study.

Some of the key properties of the population of detectable BBH, and a few differences with second-generation detectors (i.e. Advanced LIGO type instruments)  are highlighted in Ref.~\cite{VitaleDiff}.

For each of the networks above, we selected roughly 200 events, which were analyzed with the nested sampling flavor of \texttt{lalinference}~\cite{2015PhRvD..91d2003V,2010PhRvD..81f2003V}, the parameter estimation algorithm used to characterize the BBHs detected in the first science run of Advanced LIGO~\cite{GW150914-PARAMESTIM,O1BBH}.

We used a simplified precessing approximant, \texttt{IMRPhenomPv2}~\cite{Hannam:2013oca,2016PhRvD..93d4007K,2016PhRvD..93d4006H} both for simulating the signals to add in the data, and as templates for the parameter estimation algorithm.

It is worth stressing that IMRphenomPv2 does not include higher modes, while one might expect those to be relevant for high mass systems. 
While our choice is mostly driven by the lack of better waveform approximants which contain all the relevant physics and still are fast enough to compute, we can defend it by noticing that the importance of higher order modes is enhanced by large mass asymmetries~\cite{2009PhRvD..79j4023A}, while we keep the mass ratio of the simulated \sv{signals in the range $[1/3-1]$}. While this does not imply a study similar to this should not be repeated as better waveform models become available, it reassures us that the results we obtain are a good first investigation to assess the capabilities and network requirements for 3G networks.
\section{Intrinsic mass measurement}\label{Sec.IntrMass}

What is measured by GW detectors are the redshifted BH masses, from which one needs to calculate the intrinsic, astrophysically interesting, masses~\cite{Krolak1987,2005ApJ...629...15H}. The two are related by the simple relationship:

\beq
m^s = \frac{m^{det}}{1+z}
\eeq

where ``s'' stands for source frame and ``det'' for detector frame. With $m$ here we indicate any mass parameter (component masses, total mass, chirp mass \mc$\equiv \frac{(m_1\,m_2)^{3/5}}{M^{1/5}}$).
Thus, in order to get the posterior distribution for the source-frame masses one needs to estimate the redshift of the GW source.

Unfortunately, the redshift is extremely hard to measure from GW observations alone.
While it could be measurable for systems with at least one neutron star,
 if the equation of state of nuclear matter were known~\cite{2012PhRvL.108i1101M,2014PhRvX...4d1004M,2015arXiv150606590D},
 no method has been suggested to extract the redshift directly from BBH GW detections.
On the other hand, GWs provide a direct measurement of the luminosity distance to the source.
From this, the redshift can be calculated if one assumes the cosmology is known. This is the approach followed to calculate the redshifts quoted for \Event{} and \Xmas{}~\cite{O1BBH}, where the latest cosmology measured by Planck was used~\cite{2015arXiv150201589P}.

We  assume the same approach will be followed for 3G detectors, and the luminosity distance will still play a pivotal role to measure the redshift, and hence the intrinsic masses.
This could change if some intrinsic properties of black holes in BBH were discovered in the next few years that can be used to directly extract the redshift from GW measurements, however no such property seems to exist.

This is one of the first examples in which a coupling between one extrinsic parameter (distance)
 and some intrinsic parameters (masses) becomes apparent,
 while these two groups have traditionally been considered quite independent,
 in the sense that the measurement of one would not affect the other. 
Later in the life-span of advanced detectors, and even more so with 3G instruments, i.e. when cosmological distances are reached, a good estimation of the luminosity distance is paramount when measuring masses. 

\subsection{The role of polarizations}

It is well known that, within general relativity, gravitational waves have two polarizations~\cite{Maggiore}.
In the most common coordinate frame~\cite{1989JBAA...99..196H}, the luminosity distance to the source and the inclination of the orbital plane with respect to
 the line of sight enter in different ways in the two polarizations:

\begin{eqnarray}
h_+ &\propto& \frac{\left(1+\cos^2\iota\right)}{2 D_L} \\
h_\times&\propto&  \frac{\cos\iota}{D_L} 
\end{eqnarray}

Being able to measure both polarizations will thus help in breaking
 correlations and improve the estimation of the luminosity distance,
  which in turn is necessary to estimate the source-frame masses.
This is an important reason why more than one 3G detector should be built:
 measuring distances, and hence intrinsic masses,
 with a single detector would be extremely imprecise.

Having a network as large as possible is a goal already being pursued for 2G detectors.
However, in the case of 2G detectors the main driver for having more than two instruments is to reduce the uncertainty in the sky localization of the GW sources~\cite{2011PhRvD..83j2001K,VeitchMandel:2012,Fairhurst2009,VeitchMandel:2012,2011CQGra..28j5021F,2016LRR....19....1A,2014arXiv1404.5623S,2015ApJ...800...81E,2011PhRvD..84j4020V}, thus increasing the chances of identifying electromagnetic or neutrino counterparts to CBC and other sources. 
For 3G detectors, even the measurement of BH masses requires a network that can disentangle GW polarizations, since one needs to measure the luminosity distance to get the intrinsic masses~\footnote{Note that this might already be the case for 2G detectors once they reach design sensitivity}.

\section{Results}\label{Sec.Results}

In this section we report the uncertainties in the estimation of some key parameters and show how those depend on the configuration and size of the networks of Sec.~\ref{Sec.Networks}. Unless otherwise said, we will quote the 90\% credible interval divided by the true value of the parameter, and quote uncertainties in percent. Occasionally, we will report the un-normalized 90\% credible interval.

\subsection{Distance and sky position}\label{Sec.Distance}

Let us first report the uncertainties in the estimation of two important extrinsic parameters: the luminosity distance and the sky position.
As we said, both quantities will be affected by the number of detectors in the network. The effect of the numbers of detectors on sky localization uncertainties for 2G detectors has been already addressed in several papers, mostly for binary neutron stars, e.g.~\cite{VeitchMandel:2012,2011CQGra..28j5021F}. Work is ongoing to also include 3G detectors~\cite{CardiffPrep}.

While for CBC sources with one or two neutron stars the interest in their sky positions is fully justified by the fact that those systems are expected to be progenitors of short GRBs~\cite{2012ApJ...746...48M},
 there is no clear connection between BBH and EM radiation.
However, it is still interesting to report sky localization uncertainties for three reasons:
a) While unlikely, it is not impossible that BBH will in fact emit some energy in the EM, or neutrinos, as some mechanisms have been proposed~\cite{2016ApJ...819L..21L,2016ApJ...822L...9M} after the discovery of \Event{} and the alleged EM sub-threshold Fermi trigger~\cite{2016arXiv160203920C}
b) the trends we will see should be indicative of what one can expect for BNS and
c) the positions of detected BBH could be used to study the large-scale structure of the Universe~\cite{2012PhRvD..86h3512J,2016PhRvD..94b4013N} and to look for the host galaxy and calculate the cosmological parameters~\cite{2012PhRvD..86d3011D}.

\begin{figure*}[htb]
\includegraphics[width=1.9\columnwidth]{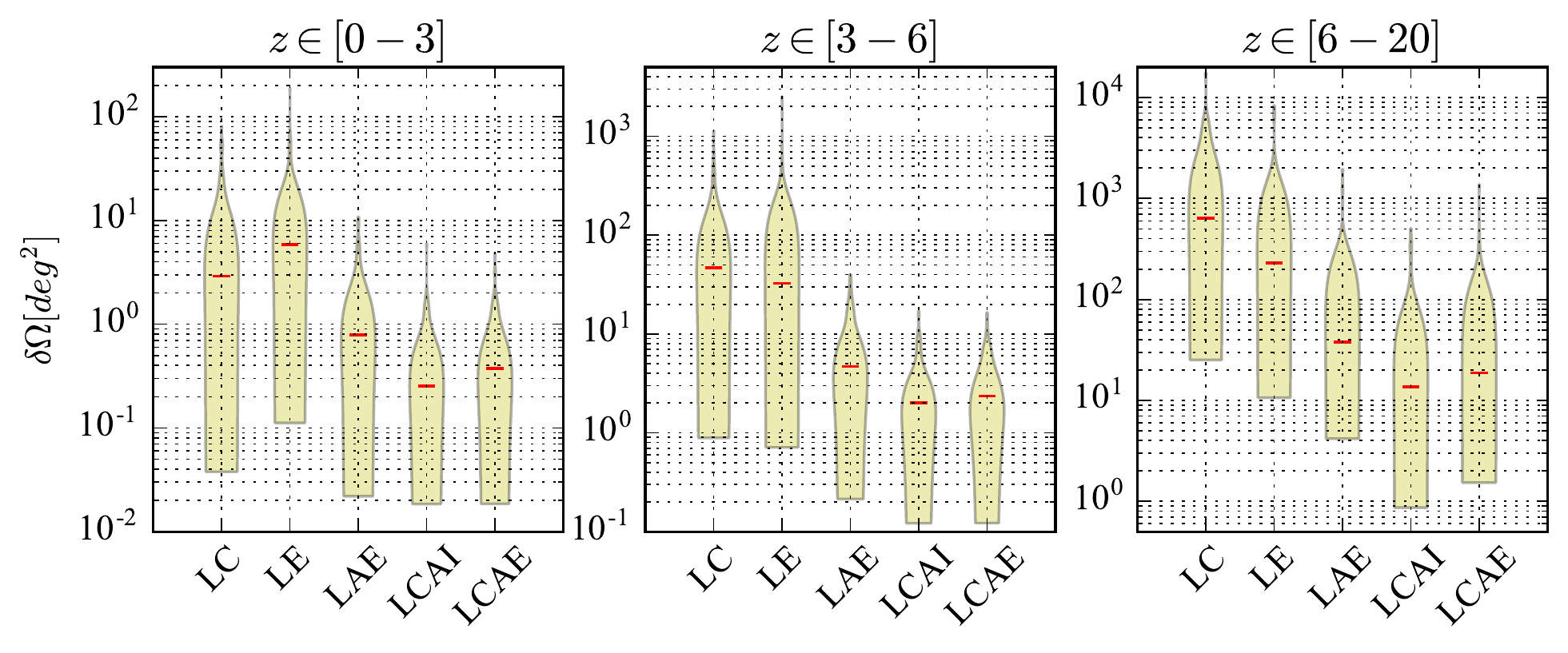}
\caption{Violin plots for the 90\% sky localization uncertainties (y axis). Each panel focuses on a different redshift bin, indicated at the top. The labels on the x axis identify the 3G networks.}
\label{Fig.Sky_violin}
\end{figure*}

With this this in mind, in Fig.~\ref{Fig.Sky_violin} we show violin plots for the 90\% credible interval for the sky position, in square degrees.  In each violin, a red horizontal line marks the median.
Each panel only uses events in a given redshift range, specified at the top, and the label in the x axis specifies the network.
The choice of redshift bins are arbitrary, and mostly chosen to ensure each bin had enough sources. 

As expected, the uncertainties are smaller for nearby sources (simply because they will on average be louder) and decrease with the number of detectors

We see how events up to redshift of 3 can be localized within 100~\degg{} even with only two 3G instruments. However for systems farther away at least three detectors are needed to keep the uncertainty below that threshold for most events. With four instruments, even events at $z>6$ can often be localized within $10$~\degg{}.
For nearby sources, localizations within 1~\degg{} would be typical for 4-instrument networks and frequent with 2-instrument networks. 
The best networks would even be able to localize a large fraction of events to within a tenths of square degree, dramatically increasing the chances of identifying eventual EM or neutrino counterparts. This, together with the small distance uncertainty for relatively nearby events (see below), will significantly reduce the number of likely host galaxies.

We also see that the improvement adding a fourth detector to the network is smaller than adding a third instrument. This has already been seen for 2G detectors~\cite{2011PhRvD..83j2001K} and can been understood as follows. Given that two polarizations must be measured, a 2-detector network is just enough. Adding a third detectors makes the problem overdetermined, which dramatically increases the polarization resolution. After that, adding a fourth detector serves mainly to increase the signal in the noise.\footnote{Ref.~\cite{2011PhRvD..83j2001K} also consider a 5-detector 2G detector network, which effectively show how the improvement plateaus after the fourth detector.}

\begin{figure*}[htb]
\includegraphics[width=1.9\columnwidth]{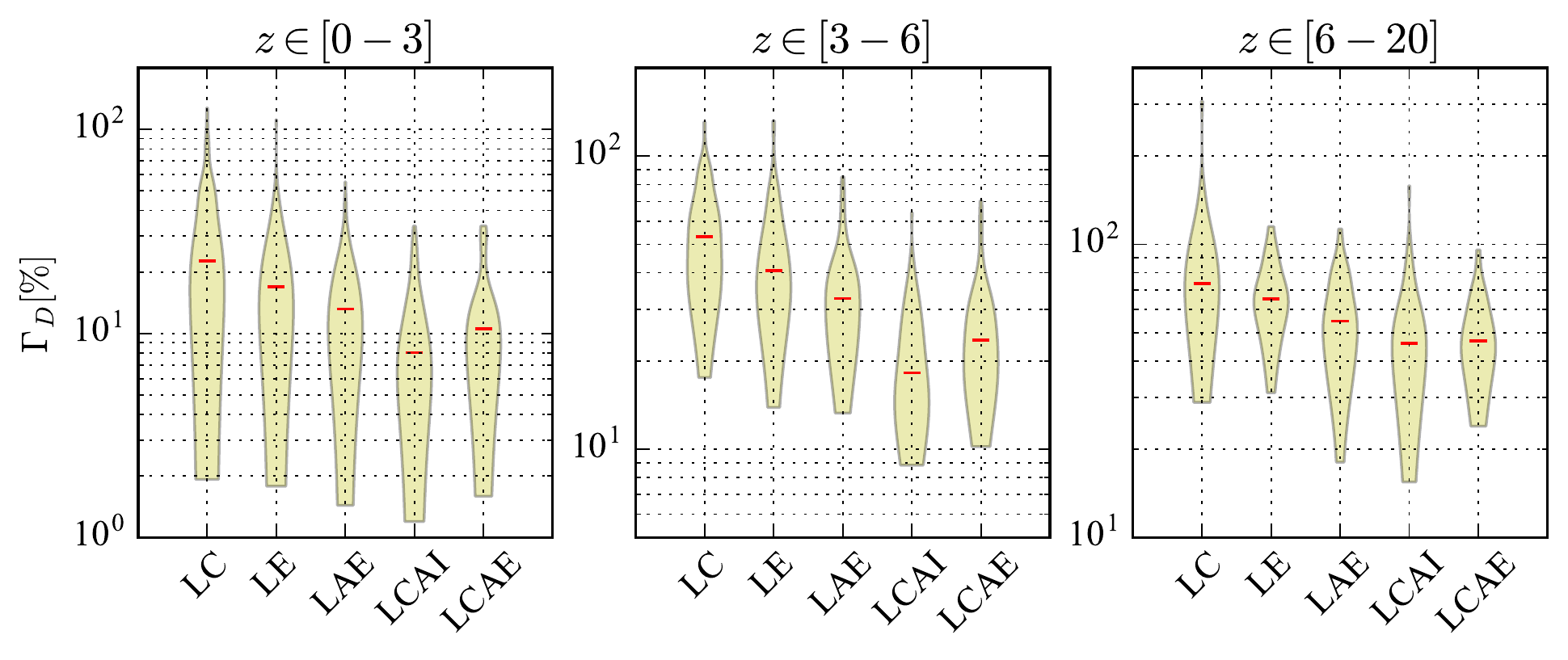}
\caption{Violin plots for the 90\% relative uncertainties for the luminosity distance (y axis). Each panel focuses on a different redshift bin, indicated at the top. The labels on the x axis identify the 3G networks.}
\label{Fig.D_violin}
\end{figure*}

In Fig.~\ref{Fig.D_violin} we present a similar plot for the 90\% credible interval relative uncertainties on the luminosity distance.
It is clear that uncertainties below 10\% will only be achieved for sources with $z\lesssim 3$. 
If it is indeed the case that sources are distributed uniform in comoving volume, i.e. with a distribution of redshift similar to the one shown in Fig.~\ref{Fig.RedshiftPrior} that peaks at $z\sim 2$, then typical uncertainties can be expected to be below 10\% for a 4-detector network, and a factor of 2 larger for 2-instrument networks.
As the redshift of the sources increase, so do the uncertainties. For sources at $z>6$, the 2-instrument networks have a significant fraction of events with relative uncertainties above 100\%, although the median stays below that value.

We notice that nearby loud sources could be characterized with extreme precision. For the redshift bin $z<3$, 10\% of events will have sky positions and distances 90\% CI uncertainties below $0.5$~\degg{} ($0.06$~\degg) and $5$\% ( $3$\%) for the 2-detector (4-detector) network, dramatically reducing the number of potential host galaxies. This will help for cosmological studies~\cite{2012PhRvD..86d3011D}.

We notice that the network LE does typically better than LC. This happens because the ET has more polarization discrimination power than a simple L-shaped detector such as the CE we assumed for the detector in China~\cite{2009CQGra..26h5012F}.
Thus, a LE network should effectively be slightly better than a network with 2 L-shaped detectors (such as LC). This is exactly what the plot confirms. We don't see the same happening for the 4-instrument networks. We explain this by noticing that since the polarization problem is already over-determined in 4-instrument networks, the small extra SNR that a CE can yield matters more than the extra polarization content of ET.

We stress that in our results we have implicitly assumed weak lensing errors can be corrected~\cite{2010CQGra..27u5006S,0004-637X-637-1-27}.
Should this not be true, weak lensing could add a systematic error to the measured luminosity distance. 
Since in this paper we are dealing with statistical errors and their dependence on the network size, and since all networks would be affected in the same way, we neglected the possibility of weak lensing errors. They should, however, be taken into account in future work as they could significantly contribute to the total (statistical plus systematic) error budget for the distance measurement at large redshifts. 

\subsection{Spins}

Before we move to the masses, we report the uncertainties on the measurement of spin parameters, another key quantity for the characterization of compact objects, and black holes in particular. In a companion paper~\cite{2016arXiv161101122V} we have reported on the measurability of spins in heavy BBH with networks of 2G detectors and found that they will be hard to measure for individual events.

\begin{figure}[htb]
\includegraphics[width=0.9\columnwidth]{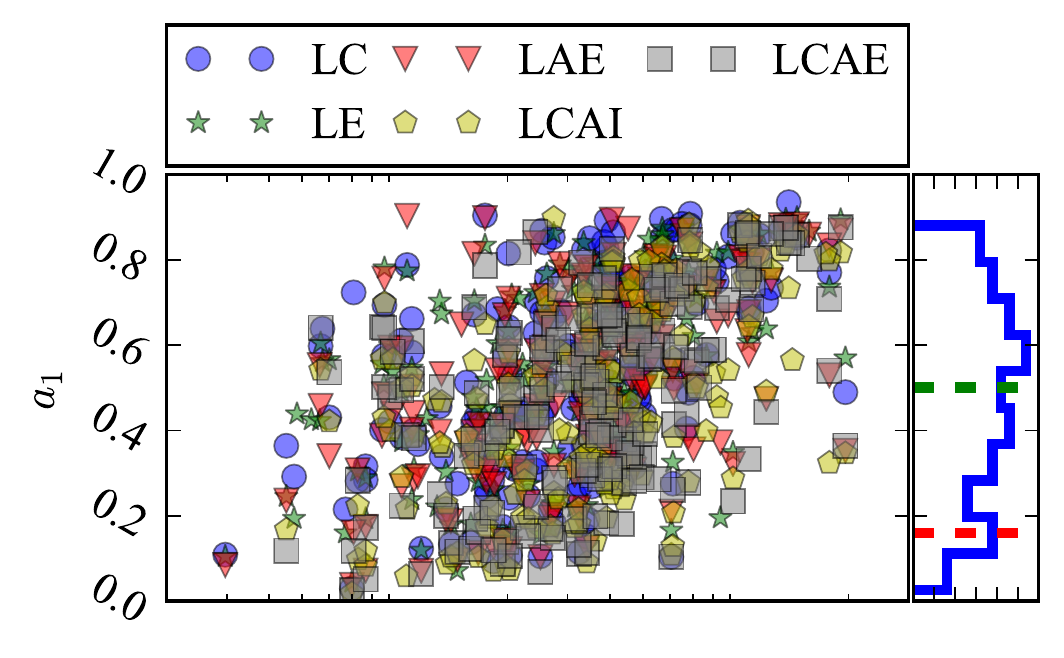}
\vskip -0.2cm
\includegraphics[width=0.9\columnwidth]{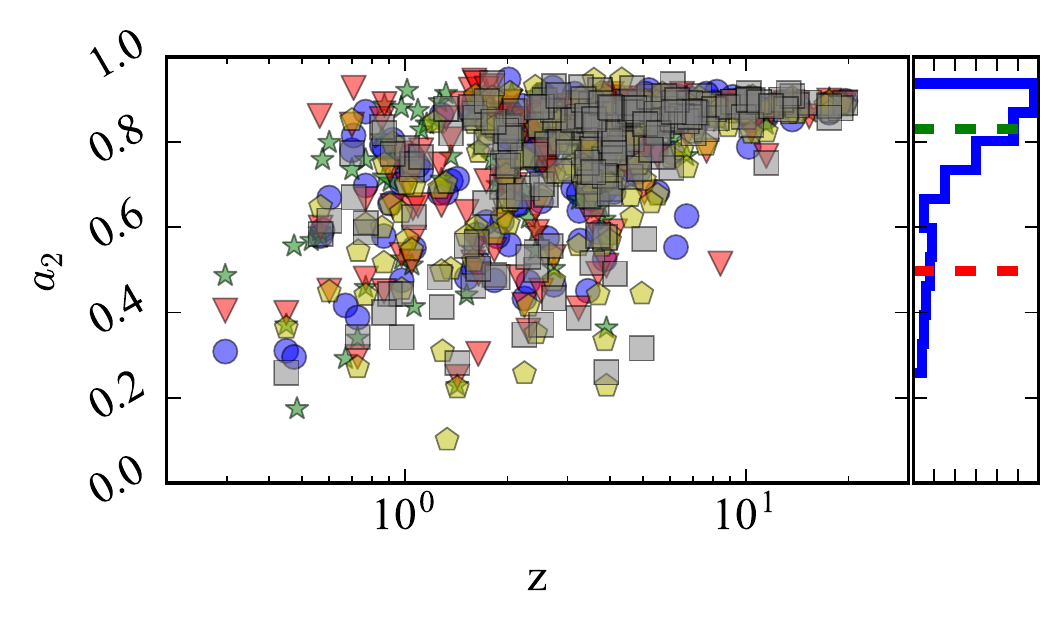}
\caption{The distribution of uncertainties for the spin magnitude of the primary (top) and secondary (bottom) black hole versus the injected redshift. All networks are show together with different symbols since they yield very similar distributions. The histogram on the right reports the distribution of the uncertainties for the LCAI network with dashed green (red) line at the 50\% (10\%) percentile.}
\label{Fig.SpinMagn}
\end{figure}

In Fig.~\ref{Fig.SpinMagn} we show the 90\% credible interval for the measurement of the spin of the primary (top) and secondary (bottom) BH for all the networks we considered, with different symbols. We see that that spin measurement is \emph{not} a strong function of the number of detectors (modulo the small extra SNR that having more detectors gives) and no clear trends are visible. The vertical histograms on the right side show the actual distribution of the uncertainties for one of the networks (LCAI, but they all look very similar). The dashed red and green line report the 10$^{th}$ and 50$^{th}$ percentiles, respectively. Comparing with similar plots for 2G detectors~\cite{2016arXiv161101122V}, we find that 3G detectors can estimate spins better. For example, for 10\% (50\%) of systems the magnitude of the primary will be estimated with uncertainties below 0.17 (0.5). For 2G detectors and BBH in the total mass range $[60-100]$\msun, the 10\% percentile is at 0.7~\cite{2016arXiv161101122V}.
The spin of the secondary is typically hard to measure, with the posterior distribution filling most of the prior for a large fraction of systems: 10\% (50\%) of the events have errors below 0.5 (0.8).

3G detectors can thus measure spins better for a population of BBH than 2G instruments. This is due to a combination of two main factors: a) most of BBH detected by 3G networks will have SNRs of several tens~\cite{VitaleDiff}; and b)  for 3G detectors, the distribution of inclinations angles will favor edge-on systems~\cite{VitaleDiff}, for which spin precession is clearly visible, if present, which reduces spin-mass correlations~\cite{2016arXiv161101122V}.

\subsection{Masses}

\begin{figure*}[htb]
\includegraphics[width=2\columnwidth]{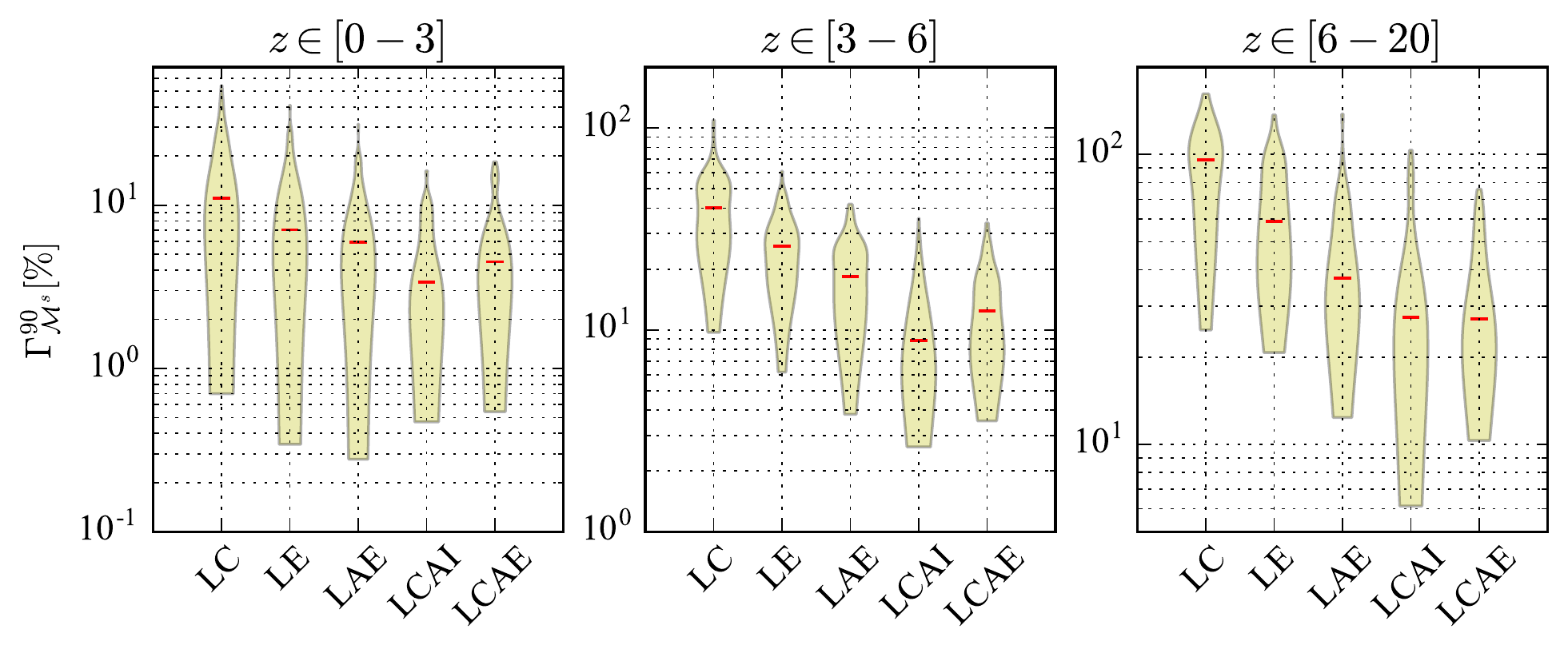}
\caption{Violin plots for the 90\% relative uncertainties for the source frame chirp mass (y axis). Each panel focuses on a different redshift bin, indicated at the top. The labels on the x axis identify the 3G networks.}
\label{Fig.Mcsource}
\end{figure*}

We now move to the main result of this study, namely the measurement of BBH intrinsic masses. 
Let us start with the chirp mass, Fig.~\ref{Fig.Mcsource}. As expected, uncertainties are lower for larger networks, whereas for the 2-detector LC network only half of the signals in the closest redshift bin have  uncertainties below the 10\% level.

The best measurements are obtained with the LCAI network, which can yield uncertainties below 10\% all the way to a redshift of several.

\begin{figure*}[htb]
\includegraphics[width=2\columnwidth]{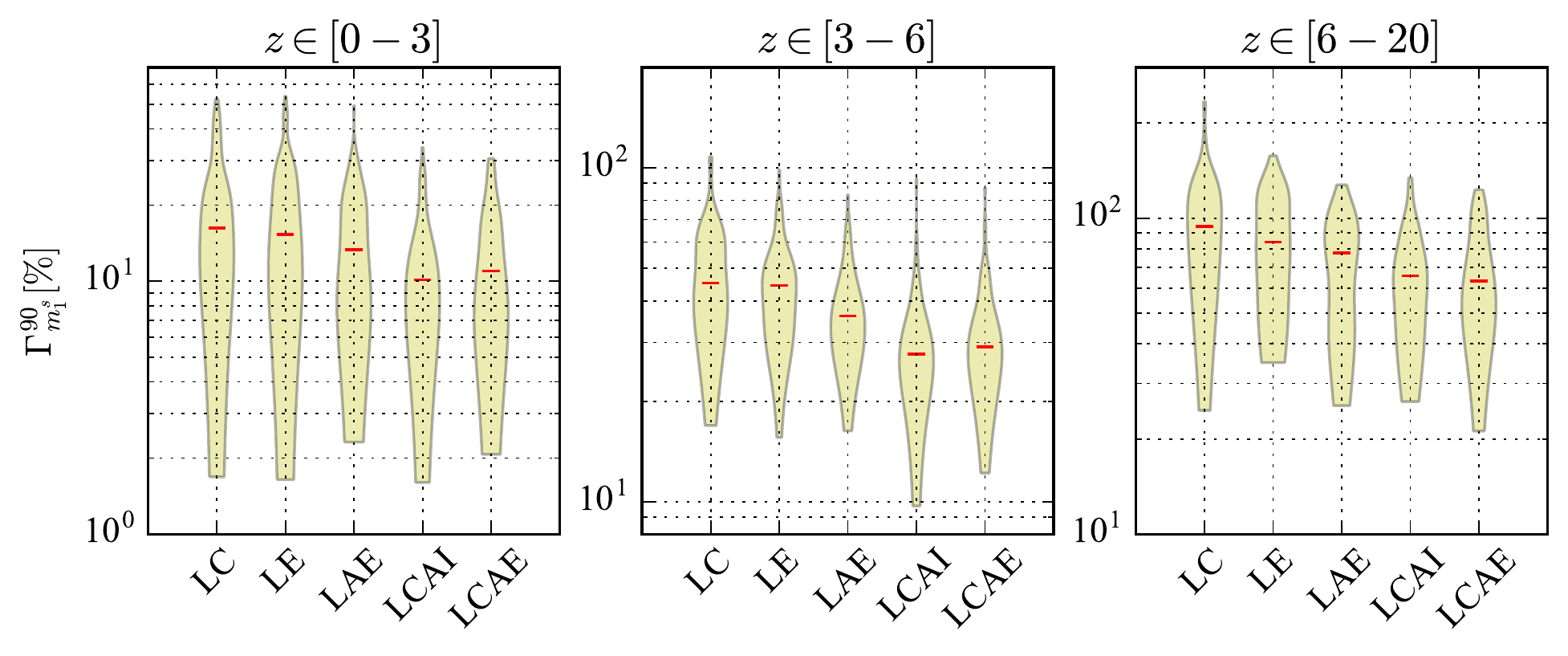}
\caption{Violin plots for the 90\% relative uncertainties for the source frame $m_1$ (y axis). Each panel focuses on a different redshift bin, indicated at the top. The labels on the x axis identify the 3G networks.}
\label{Fig.ComponentM1source}
\end{figure*}

\begin{figure*}[htb]
\includegraphics[width=2\columnwidth]{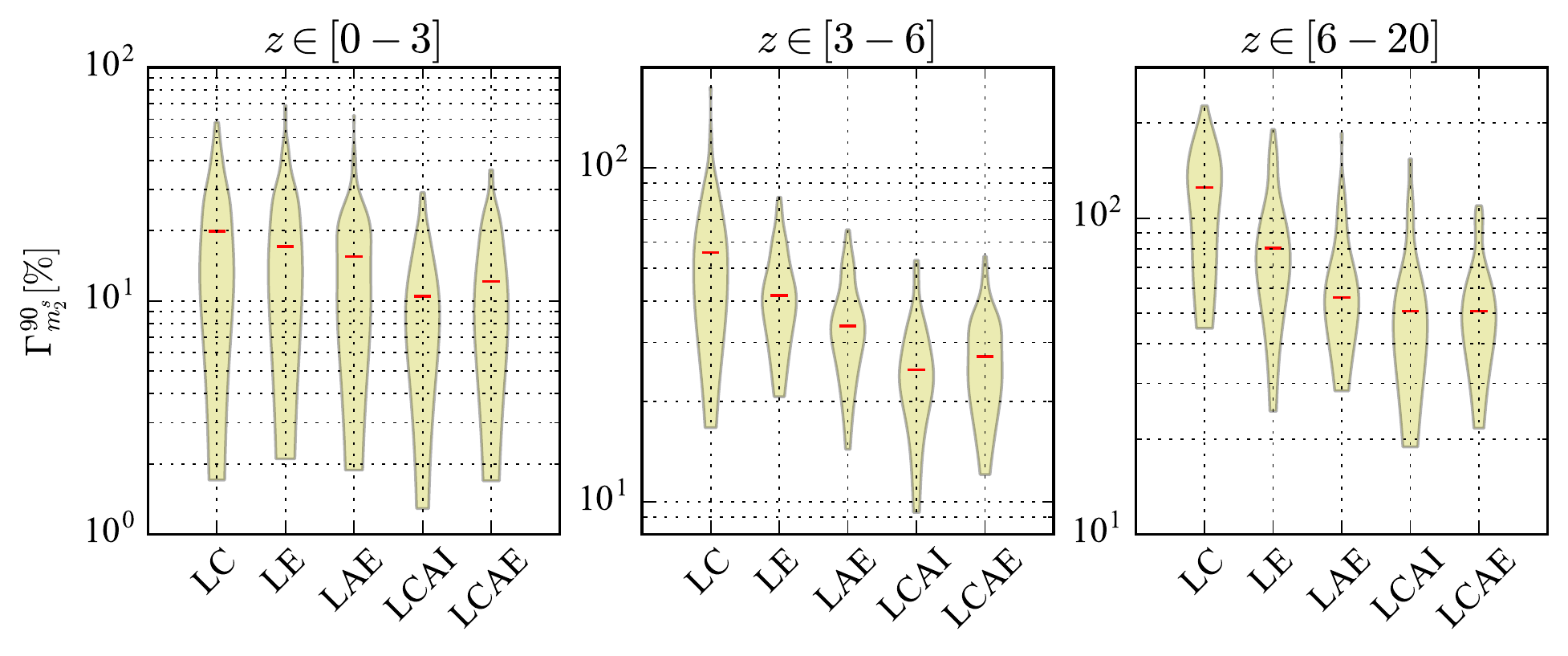}
\caption{Violin plots for the 90\% relative uncertainties for the source frame $m_2$  (y axis). Each panel focuses on a different redshift bin, indicated at the top. The labels on the x axis identify the 3G networks.}
\label{Fig.ComponentM2source}
\end{figure*}

Similar conclusions hold when considering the component masses, Figs.~\ref{Fig.ComponentM1source} and ~\ref{Fig.ComponentM2source}.
It is worth stressing that even for nearby loud events 3G networks will not typically get sub-percent uncertainties in the estimation of the component masses. For the best 10\% of sources, relative uncertainties for either component masses are below \si 5\% but larger than 1\%. This is \emph{not} due the uncertainty in the distance, but only to the fact that component masses are correlated in GW signals. Even if the redshifted masses were considered, uncertainties would still be above 1\% for all the signals we analyzed. 

In Ref.~\cite{2016PhRvL.116w1102S} it was noticed how BBH of a few tens of solar masses could be observed in both the eLISA band and in the band of ground-based detectors. Ref.~\cite{2016PhRvL.117e1102V} showed how eLISA measurements of masses (and sky position) can be used as priors in the parameter estimation analysis with a network of five 2G ground based detectors, improving e.g. the measurement of spins.
That might be less true while considering eLISA+3G detectors, since the BBH events detectable by eLISA would be at $z\lesssim0.4$~\cite{2016PhRvL.116w1102S}, for which the SNR in 3G detectors would be extremely large. In fact, component masses with 3G would be estimated at the few percent level for the closest sources, Figs.~\ref{Fig.ComponentM1source} and \ref{Fig.ComponentM2source}, comparable with what eLISA would do.

For systems in the redshift range $[0,3]$, where most events could live, given the redshift prior, the uncertainties with LC are roughly 1.5-2 times larger than what yielded by 4-instrument networks.
This ratio stays roughly the same in the other redshift bins for $m_1$ and $m_2$, while it increases for \mc.

One of the most intriguing possibilities with 3G detectors, is to detect BBH from the epoch of reionization ($z\in [6,20]$). For events at those distances, more than two instruments will be necessary to estimate both component masses with uncertainties below 100\% for most events. In that range, 4-detector networks would give typical uncertainties of only a few tens of percent.

It is worth making a final remark: given that we simulated signals uniform in comoving volume, the events at redshift of a few dominate our population, and we did not get anything closer than $z\sim0.3$. This is much higher than either \Xmas{} or \Event{}. This does not imply that nearby events will not be detected very often by 3G networks, but just that they will be detected \emph{less} often than events farther away. In a different paper, we will consider some of the research enabled by SNRs in the thousands. 


\section{Caveats}\label{Sec.Caveats}

Through this work we have made a few assumptions and choices, driven by limitations in computing power or simply by lack of better alternatives. Here we list them with the hope that this will guide future studies.

\begin{itemize}
\item We have used a waveform family without higher harmonics. While this is probably acceptable given the limited mass-ratio range we considered (the effect of higher harmonics becomes more important for large mass ratios~\cite{2009PhRvD..79j4023A}), it should be improved as fast waveforms with higher harmonics become available.
\item Given the large number of simulated sources and networks we considered in this work, to keep the computational cost manageable, we have started the analysis at 10~Hz for all detectors, while the ET would be sensitive to lower frequencies, down to a few Hz. Going down to lower frequencies could give a little SNR boost to the networks with an ET facility. We have verified that starting the analysis at 5~Hz in the ET would increase the median network SNR for our population by 9\% for the LE network, 6\% for  LAE and 2\% for LCAE. 
\item We have not considered potential systematic errors arising from weak lensing, de facto assuming lensing can be corrected. While this is acceptable when comparing the statistical performances of proposed networks, it should be taken into account if the full astrophysical capabilities of 3G networks are studied. 
\item We have used an arbitrary upper limit for the mass distribution of the sources we simulated. This is due to the lack of observational evidence, and should be revised as the advanced detectors expand our knowledge of CBCs and BHs.
\item We have assumed that BBH sources are uniformly distributed in comoving volume. Different star formation rates should be folded in, which would change the redshift distribution of the detected sources.
\end{itemize}

We plan to explore the benefit of lower starting frequencies in the Einstein Telescope in the near future. Here we just mention that a small variation in the SNR does not necessarily imply a negligible impact in the estimation of parameters. One might expect that for the sources for which the band $[\mbox{few}-10]$~Hz contains inspiral cycles that would be unaccessible otherwise, the benefit might be significant, especially for the inspiral parameters (chirp mass, mass ratio, spins).
If one assumes the black hole mass function is a steep power law which favors stellar-mass black holes~\cite{O1BBH}, most sources would have total mass of \si~20~\msun. Sources at redshifts below a few, would be redshifted to detector frame masses of no more than $100$~\msun. For these masses, inspiral cycles are already visible starting at 10~Hz.
For heavier object, or sources at larger redshifts, only the merger and ringdown would be visible starting at 10~Hz, which could make sensitivy to lower frequencies more important.

\section{Conclusions}\label{Sec.Conclusions}

The detection of GWs from two binary black hole coalescences has clearly shown that advanced ground-based detectors will detect tens of systems per year. The LIGO and Virgo detectors will reach their design sensitivity over the next few years, when stellar-mass or heavy BBH will be visible up to redshifts of roughly 1 (intermediate mass black holes would be visible up to z\si2).
New facilities will be necessary to extend the reach of ground-based detectors to redshifts of several.

R\&D is already ongoing, and several solutions for third generation detectors have been suggested.
The Einstein Telescope design consists of three 60-degree 10-Km arms interferometers  arranged to form a triangle. Built underground, it would be sensitive down to a few Hz.
The Cosmic Explorer still keeps the L-shape of existing instruments, but with increased arm-length of 40~Km. 
The sensitivity of both instruments would be over a factor of 10 better in strain than the \emph{design} sensitivity of Advanced LIGO.

Third generation detectors would thus be sensitive to BBHs up to redshifts of 10 and above, and would be able to target stellar mass or heavy black holes born from the first generation of stars.
The large horizons of these instruments would allow extremely precise measurements of mass and spins, for nearby loud events, and reconstruction of the mass evolution of BHs through cosmic history.

The estimation of masses is complicated by the fact that mass parameters are redshifted in gravitational-wave signals, so that what one measures is not the mass, but $(1+z)$ times the mass; thus a measure of the redshift must be used to convert detector-frame masses into intrinsic masses. 
However, GWs do not directly yield a measurement of the redshift, but rather of the luminosity distance, from which the redshift can be obtained if the cosmology is known. Since information about the distance is encoded in both polarizations of the GW signal, one can expect that at least two detectors are necessary to properly measure it, and hence measure the intrinsic masses.

In this paper we have shown how well some key parameters of BBH can be measured for several hypothetical networks of 3G detectors, made of two, three and four instruments.
We generated distributions of BBH with intrinsic total masses in the range \massrangeinj{} and random spins, uniformly distributed in comoving volume. 
The simulated BBH signals were then added into simulated noise of the 3G networks, and their parameters estimated using a nested sampling algorithm.

As expected, we found that the component masses and the chirp mass are estimated better as more detectors are added.
More precisely, we found that the median uncertainty is between a factor of 1.5 and  2 larger for 2-instrument networks than for 3- or 4-instrument ones.
For nearby events ($z<3$), typical 90\% credible interval uncertainties for the component masses will be around $10-20$\%, but uncertainties of a few percent will be common. For sources at large redshifts ($z>6$) more than two instruments are necessary to have median uncertainties significantly below 100\%.
Similar conclusions hold for the chirp mass.

We have verified that the estimation of spins is not affected by the network details, which is expected since spins enter the waveforms in ``redshift-free" combinations. Given that nearby events ($z<3$) can be extremely loud in 3G detectors, and that inclination angles will be isotropically distributed, precise spin estimation will be possible. Furthermore, events up to redshift of a few can be localized in the sky to within 10 \degg{} even with two instruments only, and with medians of $\sim0.3$ \degg{} if four detectors are available. This will strongly reduce the number of potential host galaxies.
We did not perform explicit simulations to assess the measurability of (eventual) deviations from general relativity. However it is clear how the possibility of accessing BBH events at SNRs of several hundreds would open new avenues for precise tests of general relativity.
This study should be updated as more realistic waveforms become available, or if significant updates are made to the design of 3G detectors. However, it already clearly shows that more than two 3G detectors should be built to maximize the science output.

\section{Acknowledgments}

Many people are thinking and working on the design and the science goals of 3G detectors.
It is a pleasure to acknowledge useful discussions with many of them. In particular the authors would like to thank R.~Adhikari, S.~Ballmer, Y.~Chen, S.~Fairhurst, A.~Freise,  B.~Sathyaprakash and D.~Sigg.

S.V. would like to thank C.~Belczynski, C.~Berry, C.~Van~Den~Broeck, T.~Dent, R.~Eisenstein, R.~Essick, W.~Farr, S.~Gaebel, T.~Regimbau and J.~Veitch for useful comments.
The authors would like to thank the referees of Phys. Rev. D for their useful comments.

The authors acknowledges the support of the National Science Foundation and the LIGO Laboratory. LIGO was constructed by the California Institute of Technology and Massachusetts Institute of Technology with funding from the National Science Foundation and operates under cooperative agreement PHY-0757058.
The authors would like to acknowledge the LIGO Data Grid clusters, without which the simulations could not have been performed. Specifically, we thank the Albert Einstein Institute in Hannover, supported by the Max-Planck-Gesellschaft, for use of the Atlas high-performance computing cluster.

This is LIGO Document P1600291.
\bibliography{draft.bib,pe.bib}
\end{document}